\documentclass[aps,prl,twocolumn,showpacs,superscriptaddress,floatfix]{revtex4}

\usepackage{graphicx}
\usepackage{amssymb}

\begin{document}

\title{Enhancement of superconductivity near the ferromagnetic quantum critical point in UCoGe}

\author{E. Slooten}
\affiliation{Van der Waals - Zeeman Institute, University of Amsterdam,
Valckenierstraat~65, 1018 XE Amsterdam, The Netherlands}
\author{T. Naka}
\affiliation{National Research Institute for Materials Science, Sengen 1-2-1,
Tsukuba, Ibaraki 305-0047, Japan}
\author{A. Gasparini}
\affiliation{Van der Waals - Zeeman Institute, University of Amsterdam,
Valckenierstraat~65, 1018 XE Amsterdam, The Netherlands}
\author{Y. K. Huang}
\affiliation{Van der Waals - Zeeman Institute, University of Amsterdam,
Valckenierstraat~65, 1018 XE Amsterdam, The Netherlands}
\author{A. de Visser}
\email{a.devisser@uva.nl} \affiliation{Van der Waals - Zeeman
Institute, University of Amsterdam, Valckenierstraat~65, 1018 XE
Amsterdam, The Netherlands}

\date{\today}

\begin{abstract}
We report a high-pressure single crystal study of the superconducting
ferromagnet UCoGe. Ac-susceptibility and resistivity measurements under
pressures up to 2.2 GPa show ferromagnetism is smoothly depressed and vanishes
at a critical pressure $p_c = 1.4$~GPa. Near the ferromagnetic critical point
superconductivity is enhanced. Upper-critical field measurements under pressure
show $B_{c2}(0)$ attains remarkably large values, which provides solid
evidence for spin-triplet superconductivity over the whole pressure range. The
obtained $p-T$ phase diagram reveals superconductivity is closely connected to
a ferromagnetic quantum critical point hidden under the superconducting `dome'.

\end{abstract}


\maketitle

The recent discovery of superconductivity in itinerant-electron ferromagnets
tuned to the border of ferromagnetic
order~\cite{Saxena-Nature-2000,Aoki-Nature-2001,Akazawa-JPCM-2004,Huy-PRL-2007}
disclosed a new research theme in the field of magnetism and superconductivity.
Notably, superconducting ferromagnets provide a unique testing
ground~\cite{Saxena-Nature-2000,Levy-Science-2005} for superconductivity not
mediated by phonons, but by magnetic interactions associated with a magnetic
quantum critical point (QCP)
~\cite{Fay-PRB-1980,Lonzarich-CUP-1997,Monthoux-Nature-2007}.
In the `traditional' model for spin-fluctuation mediated superconductivity~\cite{Fay-PRB-1980} a second-order ferromagnetic quantum phase transition takes place when the Stoner parameter
diverges, and near the critical point the exchange of longitudinal spin
fluctuations stimulates spin-triplet superconductivity. Superconductivity is
predicted to occur in the ferromagnetic as well as in the paramagnetic phase,
while at the critical point the superconducting transition temperature $T_s
\rightarrow 0$. Research into ferromagnetic superconductors will help to unravel how magnetic fluctuations can stimulate superconductivity. This novel insight might turn out to be crucial in the design of new superconducting materials.

High-pressure experiments have been instrumental in investigating
the interplay of magnetism and superconductivity. In the case of
UGe$_2$~\cite{Saxena-Nature-2000} superconductivity is found only
in the ferromagnetic phase under pressure close to the critical
point and at the critical pressure, $p_c$, ferromagnetism and
superconductivity disappear simultaneously. The ferromagnetic
transition becomes first order for $p \rightarrow p_c=
1.6$~GPa~\cite{Terashima-PRL-2001}. Moreover, a field-induced
first-order transition between two states with different
polarizations was found in the ferromagnetic
phase~\cite{Pfleiderer-PRL-2002}. Superconductivity is attributed
to critical magnetic fluctuations associated with this first order
metamagnetic transition~\cite{Sandeman-PRL-2003}, rather than with
critical spin fluctuations near $p_c$. In UIr the
ferro-to-paramagnetic phase transition remains second order under
pressure all the way to $p_c =
2.8$~GPa~\cite{Akazawa-JPCM-2004,Kobayashi-JPSJ-2007}.
Superconductivity appears in the ferromagnetic phase in a small
pressure range close to $p_c$, however, it is not observed for $p
\geq p_c$, which is at variance with the `traditional'
spin-fluctuation model~\cite{Fay-PRB-1980}. In
URhGe~\cite{Aoki-Nature-2001} ferromagnetism and superconductivity
are observed at ambient pressure. Pressure raises the Curie
temperature, $T_C$, and drives the system away from the magnetic
instability~\cite{Hardy-PhysicaB-2005}. Compelling evidence has
been presented that $p$-wave superconductivity in URhGe is
mediated by critical magnetic fluctuations associated with a
field-induced spin-reorientation process~\cite{Levy-Science-2005}.
Moreover, the same spin-reorientation process gives rise to a
remarkable re-entrant superconducting phase near the high-field
first-order quantum-critical end
point~\cite{Levy-NaturePhys-2007}.

In this Letter we show that the response to pressure of the superconducting
ferromagnet UCoGe is manifestly different. While ferromagnetism is gradually
depressed and vanishes near $p_c = 1.4$~GPa, superconductivity is enhanced and
survives in the paramagnetic phase at least up to $2.2$~GPa. The
superconducting state is remarkably robust under influence of a magnetic field,
which provides solid evidence for spin-triplet superconductivity at both sides
of $p_c$. The $p-T$ phase diagram reveals superconductivity is closely
connected to a ferromagnetic QCP hidden under the superconducting `dome'.

UCoGe crystallizes in the orthorhombic TiNiSi structure (space
group $P_{nma}$)~\cite{Canepa-JALCOM-1996}. At zero pressure UCoGe
undergoes a ferromagnetic transition at the Curie temperature
$T_{C} = 3$~K~\cite{Huy-PRL-2007}. Magnetization measurements on a
single crystal~\cite{Huy-PRL-2008} show that the ordered moment
is small, $m_{0}$= 0.07 $\mu_B$, and directed along the
orthorhombic $c$ axis. UCoGe is a  weak itinerant ferromagnet as
follows from the small  ratio of $m_{0}$ compared to the
Curie-Weiss effective moment  ($p_{eff} = 1.7 ~\mu_{B}$), and the
small value of the magnetic entropy  associated with the magnetic
phase transition~\cite{Huy-PRL-2007}. Muon-spin relaxation
measurements~\cite{DeVisser-PRL-2009} provide unambiguous proof
that weak magnetism is a bulk property which coexists with
superconductivity. For a high-quality single crystal with $RRR =
30$ ($RRR$ is the ratio of the electrical resistance  measured at
room temperature and 1 K) superconductivity is found with an onset
temperature in the resistance  $R (T)$ of $T_s ^{onset} =
0.6$~K~\cite{Huy-PRL-2008}. Specific-heat and  thermal-expansion
data~\cite{Huy-PRL-2007} were used to calculate the rates of
change of $T_C$ and $T_s$ with pressure with help of the Ehrenfest
relation for second-order phase transitions: $T_C /dp =
-2.5$~K/GPa and $dT_{s}/dp = 0.2$~K/GPa. This hinted at a
relatively low critical pressure $p_c \sim  1.2$~GPa for the
disappearance of ferromagnetic order.

\begin{figure}
\includegraphics[width=6.5cm]{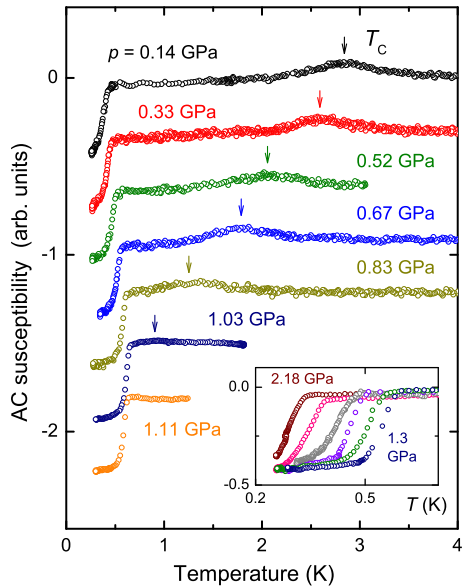}
%AC susceptibility of UCoGe plotted versus temperature at different pressures.
\caption{AC susceptibility of UCoGe plotted versus temperature at different
pressures as indicated. The curves are offset along the vertical axis for
clarity. The arrows indicate $T_C$. The inset shows $\chi_{ac}(T)$ around the
superconducting transition measured at pressures (from right to left) of 1.30,
1.52, 1.69, 1.88, 2.01 and 2.18 GPa.}
\end{figure}

The high-pressure measurements were made using a hybrid clamp cell made of
NiCrAl and CuBe alloys. Samples were mounted on a dedicated plug which was
placed in a Teflon cylinder with Daphne oil 7373 as hydrostatic pressure
transmitting medium. The pressure cell was attached to the cold plate of a
$^3$He refrigerator with a base temperature $T = 0.24$~K. The pressure was
monitored by the superconducting transition temperature of lead. The
ac-susceptibility (with frequency range $f = 113-313$~Hz and driving field
$B_{ac} \sim 10^{-6}$~T) and ac-resistivity ($f = 13$~Hz) were measured using
lock-in techniques with a low excitation current ($I = 100-300 \mu$A). The
single crystals studied were previously used in Ref.~\cite{Huy-PRL-2008}. The
ac-susceptibility was measured on a bar-shaped sample with $B_{ac} \parallel
c$~axis (long direction of the bar). For the resistivity measurements two
bar-shaped samples with the current $I$ along the $a$ axis and $c$ axis were
mounted in the pressure cell. The suppression of superconductivity by a
magnetic field was investigated by resistivity measurements in fixed magnetic
fields $B \parallel I \parallel a$ and $B \parallel I \parallel c$.

In Fig.~1 we show the temperature variation of the ac-susceptibility $\chi
_{ac}(T)$ around the magnetic and superconducting transitions at applied
pressures up to $2.18$~GPa. The Curie temperature, signalled by the maximum in
$\chi _{ac}(T)$, steadily decreases with increasing pressure up to $1.03$~GPa.
For $p \geq 1.11$~GPa the (complete) magnetic transition is no longer observed.
The bulk superconducting transition temperature, $T_{s}^{\chi}$, is identified
by the midpoint of the transition to the diamagnetic state. The width of the
superconducting transition $\Delta T_{s}^{\chi} \simeq 0.10$ K and does not
change as a function of pressure. $T_s ^{\chi}$ increases for pressures up to
$p = 1.11$~GPa, where it attains a maximum value of $0.60$~K. For larger
pressures $T_s ^{\chi}$ decreases as shown in the inset of Fig.~1. The
magnitude of the diamagnetic signal amounts to 80\% of the ideal screening
value~\cite{Huy-PRL-2007} and does not vary as a function of pressure, which
confirms superconductivity is a bulk property up to the highest pressure.

\begin{figure}
\includegraphics[width=7.5cm]{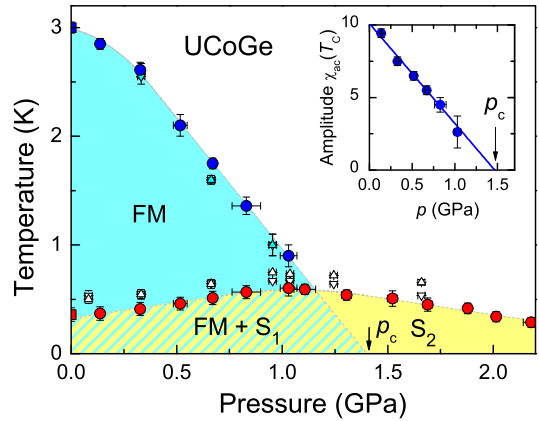}
%Pressure-temperature phase diagram of UCoGe
\caption{Pressure temperature phase diagram of UCoGe. Ferromagnetism (FM) -
blue area; superconductivity ($S_1$, $S_2$) - yellow area. $T_C (p)$
extrapolates to a ferromagnetic QCP at the critical pressure $p_c = 1.40 \pm
0.05$~GPa. Superconductivity coexists with ferromagnetism below $p_c$ -
blue-yellow hatched area. Symbols: closed blue and red circles $T_C$ and
$T_s^{\chi}$ from $\chi_{ac}(T)$; blue and white triangles $T_C$ and $T_s^{R}$
from $\rho (T)$ (up triangles $I
\parallel a$ axis, down triangles $I \parallel c$ axis); closed blue and red
squares $T_C$ and $T_s^{\chi}$ at $p = 0$ taken from a polycrystal~\cite{Huy-PRL-2007}. Inset: Amplitude of $\chi_{ac}(T)$ at $T_C$ as a
function of pressure. The data follow a linear $p-$dependence and extrapolate
to $p_c = 1.46 \pm 0.10$~GPa. }
\end{figure}

By taking the values of $T_C$ and $T_{s}^{\chi}$  from the $\chi
_{ac}(T)$ data, we construct the pressure-temperature phase
diagram shown in Fig.~2. In the diagram we have also plotted the
magnetic and superconducting transition temperatures extracted
from resistivity data, $\rho (T)$, taken in a separate
high-pressure run. Here $T_C$ is identified by a kink in $\rho
(T)$~\cite{Huy-PRL-2008} and the superconducting transition
temperature, $T_{s}^{R}$, is taken as the midpoint to the zero
resistance state. The width of the superconducting transition
$\Delta T_{s}^{R} \leq 0.10$ K at all pressures. The values of
$T_{s}^{R}$ systematically exceed $T_{s}^{\chi}$, in agreement
with the earlier observation~\cite{Huy-PRL-2007} that the
diamagnetic signal appears when the resistive transition is
complete. For $p \gtrsim 0.4$~GPa $T_C$ follows a linear
suppression at a rate of $-2.4$~K/GPa. The phase line $T_C$(p)
intersects the superconducting phase boundary near $p \approx
1.16$~GPa and its linear extrapolation yields the critical
pressure for the suppression of ferromagnetic order $p_{c} = 1.40
\pm 0.05$~GPa. An almost equal value $p_c = 1.46 \pm 0.10$~GPa is
deduced from the pressure variation of the amplitude of $\chi
_{ac}$ at $T_C$ (inset Fig.~2). The data reveal the magnetic
transition is continuous over the entire pressure range -
hysteresis in the magnetic signal is absent - and consequently
ferromagnetic order vanishes at a second-order quantum-critical
point. As we can not detect the weak $\chi_{ac}$ signal at $T_C$
in the superconducting phase, the possibility that $T_C$ vanishes
abruptly near $p = 1.1$~GPa cannot be ruled out completely. In
this case the nature of the phase transition becomes first order
and the phase line $T_C (p)$ terminates at a first order quantum
end point~\cite{Belitz-PRL-1999}.

The high-pressure data irrevocably demonstrate that
superconductivity persists in the paramagnetic regime. This marks
a pronounced difference with the other superconducting
ferromagnets, in which superconductivity is confined to the
ferromagnetic state. The phase diagram shown in Fig.~2 clarifies a
comparable diagram obtained by a high-pressure transport study on
a polycrystalline sample~\cite{Hassinger-JPSJ-2008} with
relatively broad magnetic and superconducting transitions, which
hampered notably a proper determination of $T_C (p)$. We identify
the characteristic pressure $p^* \approx 0.8$~GPa at which a
change in the superconducting properties was
reported~\cite{Hassinger-JPSJ-2008} as the pressure $p \approx
1.16$~GPa at which $T_C$ drops below $T_s$.

The presence of a second-order ferromagnetic QCP under the
superconducting dome may provide an essential clue for
understanding superconductivity in UCoGe. The phase diagram of
UCoGe, reported in Fig. 2, does not obey the `traditional' spin
fluctuation scenario~\cite{Fay-PRB-1980}, as $T_s$ remains finite
at $p_c$. Clearly, more sophisticated
models~\cite{Roussev-PRB-2001,Monthoux-PRB-2001,Belitz-PRB-2004}
are needed. Notice, a non-zero transition temperature at $p_c$ was
obtained in Ref.~\cite{Roussev-PRB-2001}, where the spin-triplet
pairing strength in the paramagnetic regime was investigated using
an Eliashberg treatment and imposing a strong Ising anisotropy of
the critical fluctuations.

UCoGe is unique because it is the only superconducting ferromagnet
which has a superconducting phase in the paramagnetic regime.
Symmetry-group considerations~\cite{Mineev-PRB-2004} tell us that
this state (which we label $S_2$ differs from the superconducting
state (label $S_1$) in the ferromagnetic phase. In the
paramagnetic state ($T > T_C > T_s$) the symmetry group is given
by $G^{sym} = G \times  T \times  U(1)$, where $G$ represents the
point-group symmetry of the orthorhombic lattice, $T$ denotes
time-reversal symmetry and $U(1)$ is gauge symmetry. Below $T_C$
time-reversal symmetry is broken, and in the superconducting phase
$S_1$ gauge symmetry is broken as well. However, in the
paramagnetic superconducting regime the phase $S_2$ breaks only
gauge symmetry. For uniaxial ferromagnets with orthorhombic
crystal symmetry the possible unconventional superconducting
states in the presence of strong spin-orbit coupling have been
worked out in detail~\cite{Mineev-PRB-2004}, and can be
discriminated in close analogy with the familiar superfluid phases
of $^3$He~\cite{Mineev-arXiv-2008}. In the ferromagnetic phase
exchange splitting results in spin-up and spin-down  Fermi
surfaces. At both Fermi surfaces an attractive pairing potential
may give rise to Cooper pairs, which form from electrons with
opposite momentum and necessarily have the same spin.
Consequently, a superconducting ferromagnet is essentially a
non-unitary two-band
superconductor~\cite{Mineev-PRB-2004,Mineev-arXiv-2008} with
equal-spin pairing Cooper states $\mid \uparrow \uparrow \rangle $
and $\mid \downarrow \downarrow \rangle$. Below $T_C$, the
superconducting phase is similar to the $A_2$ phase of liquid
$^3$He (i.e. the $A$ phase in field), which is a linear
combination of the Cooper pair states $\mid S_z = 1, m = 1
\rangle$ and $\mid S_z = -1, m = 1 \rangle$ with different
populations~\cite{Mineev-arXiv-2008}. In the paramagnetic state
the degeneracy of the spin-up and spin-down band is restored, as
well as time-reversal symmetry. There are two candidate
states~\cite{Mineev-arXiv-2008} for $S_2$: ({\it i}) a
conventional spin-singlet state, or ({\it ii}) a unitary
spin-triplet state, like the planar state of superfluid $^3$He,
which is an equally weighted superposition of the two states with
$\mid S_z = 1, m = -1 \rangle$ and $\mid S_z = -1, m = 1 \rangle$.
In the following paragraph we present measurements of the upper
critical field $B_{c2}$ under pressure, which provide evidence for
the latter scenario.

\begin{figure}
\includegraphics[width=7.5cm]{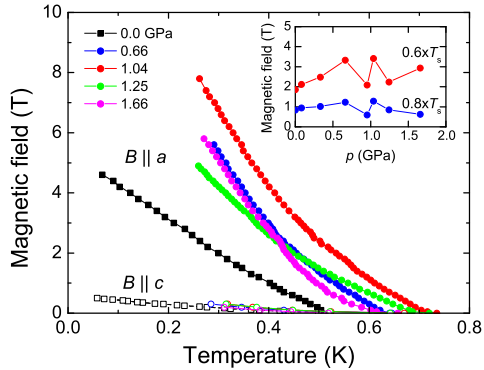}
%Temperature variation of the upper critical field at different pressures
\caption{Temperature variation of the upper critical field of UCoGe at
different pressures. Data points are collected from resistance data, measured
in fixed fields $B \parallel a$ (closed symbols) and $B \parallel c$ (open
symbols). The error bars are given by the size of the symbols. $T_s ^R (B)$ is
taken at the midpoint of the transition to $R=0$. Data for this sample at $p =
0$ are taken from Ref.~\cite{Huy-PRL-2008}. For $B
\parallel a$, $B_{c2}$ shows a strong enhancement on approach of the critical
pressure $p_c$ (blue, red and green curves). $B_{c2}$ remains
large at 1.66~GPa in the paramagnetic phase (magenta curve). For
$B
\parallel c$ the data taken at various pressures almost overlap. Inset: $B_{c2} (p)$ at $0.6$ and
$0.8 \times T_s$ after normalization $T_s (p)/T_s (p = 0) = 1$ for $B=0$. The
error bars are given by the size of the symbols.}
\end{figure}

The suppression of superconductivity by a magnetic field was investigated at
selected pressures by electrical resistivity measurements in fixed magnetic
fields. The results are reported in Fig.~3. The upper-critical field for $B
\parallel a$, $B_{c2} ^a$, shows a remarkable enhancement on approach of the critical
pressure $p_c$. By comparing with the temperature variation
$B_{c2} ^a (T)$ at ambient pressure~\cite{Huy-PRL-2008} we arrive
at a conservative estimate for $B_{c2} ^a (T \rightarrow 0)$ of 15
T near $p_c$, which exceeds the ambient pressure value by a factor
3. $B_{c2} ^a$ remains large in the whole pressure range as
demonstrated in the inset of Fig.~3, where the critical field at
0.6 and 0.8$\times$ the reduced temperature $T_s$ is plotted
versus pressure. The large values of $B_{c2} ^a (T \rightarrow 0)$
provide solid evidence for spin-triplet Cooper pairing, as the
Pauli paramagnetic limit for spin-singlet superconductivity
$B_{c2}^{Pauli} \approx 1.83 \times T_s = 1.3$~T. Thus the data
taken in the paramagnetic phase at $p=1.66$~GPa lead to the
important conclusion that spin-triplet superconductivity, most
likely with a $^3$He-like planar-state~\cite{Mineev-arXiv-2008},
persists in the paramagnetic phase. The anisotropy $B_{c2} ^a \gg
B_{c2} ^c$ yields support~\cite{Huy-PRL-2008} for an axial
superconducting gap function with point nodes along the direction
($c$ axis) of the ordered moment $m_0$. This anisotropy is
preserved under pressure.

Finally, we turn to the unusual upward curvature in $B_{c2} ^a
(T)$. For 3D materials an upward curvature in $B_{c2}(T)$ is
normally a signature of two (or multiple)-band
superconductivity~\cite{Shulga-PRL-1998}. For a two-band
ferromagnetic superconductor it can naturally be
attributed~\cite{Huy-PRL-2008,Mineev-PRB-2004} to the
field-induced population redistribution of the states $\mid
\uparrow \uparrow \rangle $ and $\mid \downarrow \downarrow
\rangle$. This calls for theoretical calculations of $B_{c2}$
using a linearized Ginzburg-Landau equation including gradient
terms~\cite{Mineev-PRB-2004} for the orthorhombic symmetry case.
The intricate non-monotonic variation of $T_s$ with field and
pressure is possibly related to a rapid variation of the different
anisotropy coefficients of the gradient terms near the critical
point. We stress, however, that the observed variation in $B_{c2}
(T,p)$ is an intrinsic property and largely exceeds the
experimental uncertainty. Another interesting scenario for the
upward curvature in $B_{c2}$ is the proximity to a field-induced
QCP, like reported~\cite{Levy-Science-2005,Levy-NaturePhys-2007}
for URhGe. However, high-field magnetoresistance data  taken at
ambient pressure~\cite{Gasparini-unpublished-2009} have not
revealed a metamagnetic transition so far ($B \leq 30$~T).

In conclusion, we have measured the $p-T$ phase diagram of the superconducting
ferromagnet UCoGe for high-quality single crystals. Ferromagnetism is smoothly
depressed and vanishes at the critical pressure $p_c = 1.4$~GPa. Near the
ferromagnetic critical point $T_s$ goes through a maximum and superconductivity
is enhanced. The upper critical field $B_{c2}(T)$ shows an unusual upward
curvature, which extrapolates to remarkably large values of $B_{c2}(0)$ at both
sides of $p_c$. This provides solid evidence for spin-triplet superconductivity
over the whole pressure range. The obtained $p-T$ phase diagram is manifestly
different from that of other superconducting ferromagnets, notably because
superconductivity is enhanced near $p_c$ and persist in the paramagnetic phase.
This reveals superconductivity is closely connected to a ferromagnetic QCP
hidden under the superconducting `dome'. We conclude the $p-T-B$ phase diagram
of UCoGe provides a unique opportunity to explore unconventional
superconductivity stimulated by magnetic interactions.

This work was part of the research program of FOM (Dutch
Foundation for the Fundamental Research of Matter). T. Naka is
grateful to NWO (Dutch Organisation for Scientific Research) for
receiving a visitor grant.

\end{document}